\documentclass[prl,twocolumn]{revtex4-1}

\usepackage{times}

\usepackage{graphicx,float}
\usepackage[normalem]{ulem}
\usepackage{color}
\begin{document}

\title{Quantum sensing of local magnetic field texture in strongly correlated electron systems under extreme conditions}

\author{King Yau Yip$^{1,\S}$}
\author{Kin On Ho$^{1,\S}$}
\author{King Yiu Yu$^{1,\S}$}
\author{Yang Chen$^1$}
\author{Wei Zhang$^1$}
\author{S. Kasahara$^2$}
\author{Y. Mizukami$^{2,3}$}
\author{T. Shibauchi$^{2,3}$}
\author{Y. Matsuda$^2$}
\author{Swee K. Goh$^{1,4}$}
\email{skgoh@cuhk.edu.hk}
\author{Sen Yang$^{1,4}$}
\email{syang@cuhk.edu.hk}

\affiliation{$^{\S}$These authors contributed equally to this work}
\affiliation{$^1$Department of Physics, The Chinese University of Hong Kong, Shatin, New Territories, Hong Kong, China}
\affiliation{$^2$Department of Physics, Kyoto University, Kyoto 606-8502, Japan}
\affiliation{$^3$Department of Advanced Materials Science, University of Tokyo, Kashiwa 277-8561, Japan}
\affiliation{$^4$Shenzhen Research Institute, The Chinese University of Hong Kong, Shatin, New Territories, Hong Kong, China}



\begin{abstract}
An important feature of strong correlated electron systems is the tunability between interesting ground states such as
unconventional superconductivity and exotic magnetism. 
Pressure is a clean, continuous and systematic tuning parameter. However, due to the restricted accessibility introduced by high-pressure devices,
compatible magnetic field sensors with sufficient sensitivity are rare. This greatly limits the detections and detailed studies of pressure-induced phenomena. Here, we utilize nitrogen vacancy (NV) centers in diamond as a powerful, spatially-resolved vector field sensor for material research under pressure at cryogenic temperatures. Using a single crystal of BaFe$_2$(As$_{0.59}$P$_{0.41}$)$_2$ as an example, we extract the superconducting transition temperature ($T_c$), the local magnetic field profile in the Meissner state and the critical fields ($H_{c1}$ and $H_{c2}$). The method developed in this work will become a unique tool for tuning, probing and understanding quantum many body systems.
\end{abstract}

\maketitle

Strongly correlated electronic systems are rich playgrounds to realize a wide variety of phases. Due to the large degrees of freedom inherent in this class of materials, these phases are naturally sensitive to external perturbations. 
Take superconductivity for example, Bardeen-Cooper-Schrieffer theory explicitly highlights the relevance of the superconducting transition temperature $T_{\rm c}$ to both the interaction strength and the density of states at the Fermi level \cite{Tinkhambook}. Moreover, superconductivity is frequently found to compete with other phases, including magnetically, structurally, and electronically ordered states ({\it e.g.} Refs. \cite{Mathur1998, Saxena2000, Goh2015, Hosoi2016}). Therefore, the ability to subject the material system to suitable tuning parameters is the major experimental tool for reaching novel phases.

One of the most successful tuning parameters is hydrostatic pressure, which changes the electronic structure and the interaction strength without introducing additional chemical inhomogeneity to the sample. Moreover, for many systems, pressure is the only way to reach certain quantum states. Pressure has played an influential role on stabilizing superconductivity by suppressing the competing phases. For example, in the heavy fermion intermetallic CePd$_2$Si$_2$, pressure suppresses the antiferromagnetic state and induces a superconducting phase with a $T_{\rm c}$ that peaks at $\sim$~28~kbar \cite{Mathur1998}. In iron-based system BaFe$_2$As$_2$, superconductivity can similarly be induced by pressure upon the suppression of the spin-density wave state ~\cite{Paglione2010}. More recently, superconducting state with a remarkably high $T_{\rm c}$ of 203~K has been reported in H$_{3}$S pressurized to 155~GPa ~\cite{Eremets-Nature}. These results not only reinforce the view that pressure is a powerful tuning parameter, but also call for the need to study the microscopic details of the superconductivity under extremely high pressure.

To generate high pressure, the sample is enclosed in a pressure cell that is orders-of-magnitude larger than the sample itself. Moreover, to ensure a stable pressure environment, electrical accessibility to the sample volume is severely restricted. Cryogenic conditions place further restrictions. 
Under these demanding experimental conditions, very few detection methods can be applied. Especially, having a robust DC magnetic field sensor in the immediate vicinity of the sample is a major experimental challenge, which we overcome in this work successfully.

Negatively charged NV center is a point defect in diamond with a spin-1 ground state. Due to its spin dependent fluorescence rate, electron spin resonance (ESR) spectrum can be measured via optically detected magnetic resonance (ODMR) method. From these spectra, we can derive the magnetic field with sensitivity of $\rm \mu T$ $\rm Hz^{-1/2}$ ~\cite{Wrachtrup06,Hollenberg13,Wrachtrup08,Lukin08,Vincent_review14}, as well as electrical field, temperature and mechanical strain ~\cite{Wrachtrup13, Lukin13, Doherty14,Budker14,Prozorov18,Prozorov_arxiv} (see Supplementary). Most importantly, NV centers can sense both the magnitude and the direction of the field, and it can work  up to 70~GPa ~\cite{Doherty14, Curro17}.
Furthermore, due to its small size, NV center naturally provides high spatial resolution, making the microscopic study of quantum many body features possible.
With these motivations in mind, we combine the field sensing capability of nitrogen-vacancy centers and the optical accessibility of a Moissanite anvil cell to probe the local magnetic field configuration around the sample at high pressures.
In this work, we demonstrate its potential by directly probing the Meissner effect in a type~II superconductor BaFe$_2$(As$_{0.59}$P$_{0.41}$)$_2$ under pressure.

The sample we use is a piece of single crystalline BaFe$_2$(As$_{1 - x}$P$_x$)$_2$ with $x=0.41$, which comes from the ultraclean family BaFe$_2$(As$_{1 - x}$P$_x$)$_2$ ~\cite{kasahara2010}. At $x=0.33$, $T_c$ is maximized and displays a clear evidence of a quantum critical point ~\cite{Matsuda12,kasahara2014}. Hence, BaFe$_2$(As$_{1 - x}$P$_x$)$_2$ is an ideal platform to explore the interplay between superconductivity and quantum criticality.

\begin{figure}
\center
\includegraphics[width=0.5\textwidth]{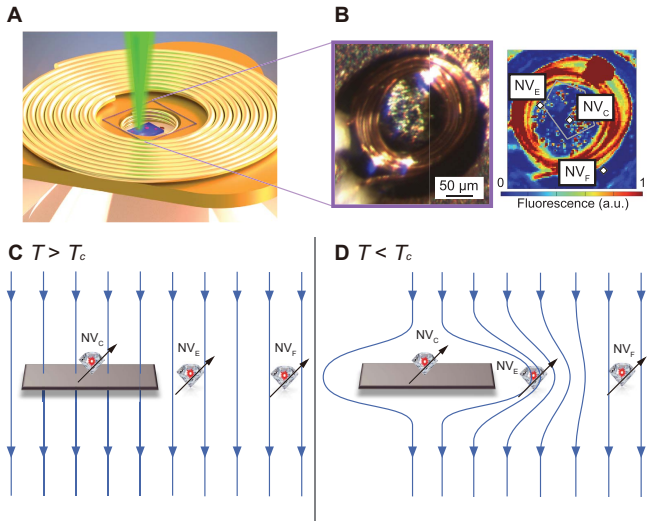}
	\caption{\textbf{Schematic illustration of experimental configurations and detection concepts.}
   (\textbf{a}) An illustration of our pressure cell design. The sample (shown in blue) is located in the high pressure chamber together with a collection of diamond particles (shown as red spots). Each diamond particle is a sensitive local field sensor. The laser is directed towards the high pressure chamber through the top moissanite anvil. The microwave is provided by a miniature microcoil in the close proximity of the sample, allowing an efficient transmission of microwave power.
The larger coil is added to serve as the modulation coil for auxiliary AC susceptibility measurements \cite{Alireza,Goh17} (see Supplementary). The metal part beneath the modulation coil is the gasket. (\textbf{b}) (\textit{left}) Photograph of the microcoil with sample on top of the anvil, and (\textit{right}), fluorescence image from the confocal scan showing the microcoil and NV centers. The shape of the sample is traced by the pentagon. The location of three particular diamond particles ${\rm NV_C}$, ${\rm NV_E}$, and ${\rm NV_F}$ are marked. ${\rm NV_C}$ is near the center of the top surface, ${\rm NV_E}$ is near the edge, and ${\rm NV_F}$ is far away from sample and serves as a control sensor. The fluorescence is collected between 650~nm and 800~nm. (\textbf{c, d}) Magnetic field profile around the sample under a weak applied magnetic field when $T > T_c$ (\textbf{c}) and $T < T_c$ (\textbf{d}). The expulsion of the magnetic field when $T < T_c$ is the Meissner effect. The alteration of the field profile in the presence of the superconductor provides an ideal platform to demonstrate the performance of our sensor to probe the complete field vectors with spatially resolution under pressure.}
\label{fig:1}
\end{figure}

Fig. 1(a) is a schematic illustration of the interior of our cell setup, showing the relative positions of the sample and the excitation microcoil as well as the direction of the laser beam. Figure 1B shows the photograph and the fluorescence image in the immediate vicinity of the sample. The close proximity of the microcoil and the sample ensures the efficient transmission of the microwave power to the sample space, where NV centers are located. Bright spots in fluorescence image are from NV centers in diamond particles, which are spread on the sample surface and mixed with pressure transmitting fluid. The typical size of diamond particles (1$\mu$m) is chosen to be smaller than the optical resolution for better sensitivity, but bigger than the vortex lattice constant $a_V$ (discussion in Supplementary). In this work, three diamond particles are chosen strategically:  ${\rm NV_C}$ is near the center of the sample, ${\rm NV_E}$ is off the side near the edge of the sample, and ${\rm NV_F}$ is far away from the sample.

Under a weak external magnetic field, for $T > T_c$, the sample is in the normal state and the magnetic field felt by NV centers is the same as the external magnetic field (Fig. 1(c)). However, for $T<T_c$, the expulsion of the magnetic field from the sample (Meissner effect) alters the field profile near the surface of the material, which can be felt by NV centers on the sample surface: for ${\rm NV_C}$, the effective magnetic field is greatly reduced; for ${\rm NV_E}$, the effective magnetic field is greatly enhanced (Fig. 1(d)). Additionally, for a type-II superconductor, when the applied field is higher than a threshold value ($H_{c1}$), the field begins to thread through the sample, resulting in a vortex state. The vortex state can be completely destroyed at a higher threshold field ($H_{c2}$), at which the superconductor returns to the normal state. All these in-field behaviors can be profiled by the NV centers located right on the sample surface.

\begin{figure*}
\includegraphics[width=1.0\textwidth]{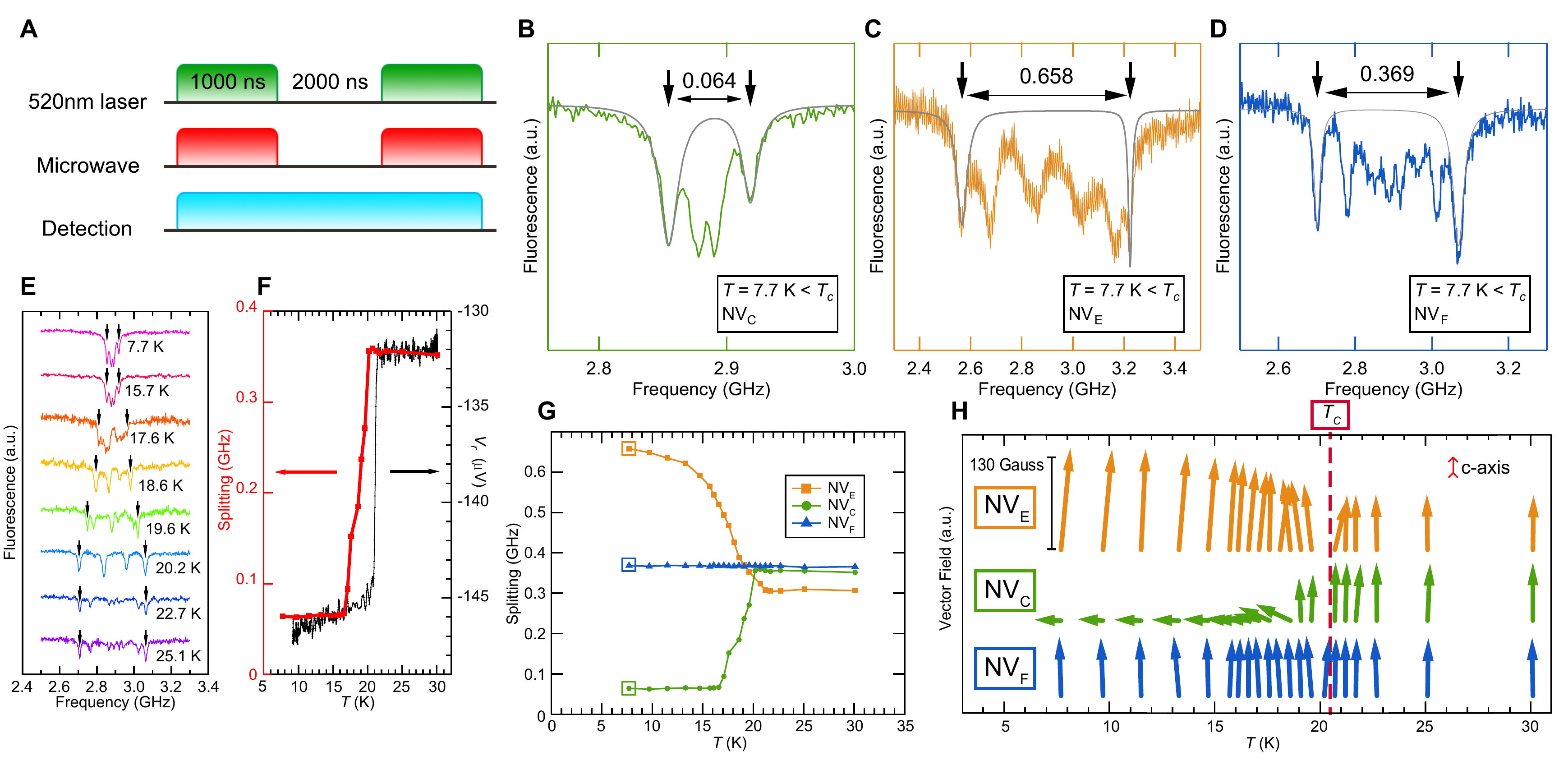}
	\caption{\textbf{The Meissner effect sensed by NV centers at 8.3~kbar.}
   (\textbf{a}) Pulse sequence used for ODMR measurements. (\textbf{b-d}) ODMR spectra of each diamond particle at 7.7~K. The Lorentzian fits for determining the Zeeman splitting are marked with grey lines. (\textbf{e}) ODMR spectra of NV centers in ${\rm NV_C}$ at different temperatures. (\textbf{f}) Comparison between the ODMR method (red) and AC susceptibility method (black) in determining the transition temperature $T_c$. 
(\textbf{g}) The change of the Zeeman splitting for NV centers in ${\rm NV_C}$, ${\rm NV_E}$, and ${\rm NV_F}$ as a function of temperature. (\textbf{h}) The variation of the local magnetic field vectors felt by NV centers in ${\rm NV_C}$, ${\rm NV_E}$, and ${\rm NV_F}$ across the superconducting phase transition. The vertical direction is the $c$-axis of the sample. The ODMR measurements were conducted with a laser power of 10~$\mu$W and a peak microwave power of 30~mW. An external $B$ field of 68~G is applied along the $z$-axis, which is parallel to the $c$-axis of the sample.}
\label{fig:2}
\end{figure*}

The Meissner effect is measured with the pulse sequence shown in Fig. 2(a). To avoid heating caused by microwaves and laser irradiation, we devise a measurement protocol to mitigate measurement-induced perturbations to the superconducting state (details in Supplementary). Fig. 2(b-d) displays representative ODMR spectra at 8.3~kbar from these three diamond particles, when the sample temperature ($\sim$7.7~K) is much lower than $T_c$ ($\sim$ 20.4~K at 8.3~kbar, determined using AC susceptibility at zero field, see Supplementary). The ODMR spectra show different splittings. Zeeman splitting of ${\rm NV_C}$ ($\sim$64~MHz) is 10 times smaller than that of ${\rm NV_E}$ ($\sim$658~MHz) when $T<T_c$ while the difference is much smaller when $T>T_c$. Fig. 2(e) shows the ODMR spectra of ${\rm NV_C}$ at different temperatures, from which the splitting can be extracted and plotted in Figure 2F. Upon warming up, the degree of splitting remains nearly constant initially, before experiencing a drastic increase that sets in at around 17~K. Above 21~K, the splitting levels off again. To demonstrate the relevance of this feature to superconductivity, we additionally collect the AC susceptibility data in the same experiment, which is possible because of the additional modulation coil added to our experimental configuration. Using the microcoil as the pickup coil, a sharp drop in the AC susceptibility, signifying the superconducting transition \cite{Lina10, Goh17}, is detected at the same temperature (Figure 2F). Both methods agree well on the measurement of $T_c$.

The change of the local magnetic field distribution can also be seen in the temperature evolution of the splitting for ${\rm NV_E}$ and ${\rm NV_F}$, shown in Fig. 2(g). Contrary to the behavior of ${\rm NV_C}$, the degree of splitting decreases for ${\rm NV_E}$ upon warming up. As a reference, the splitting for ${\rm NV_F}$ is nearly constant in temperature, which can be understood because ${\rm NV_F}$ is far from the superconductor so that the total field does not change. These observations agree nicely with the expectation from the Meissner effect as explained earlier. There is no significant change of linewidth or the overall contrast of the ESR lines. This can be understood since the diamond particles are much smaller in size compared with the magnetic field gradient induced by Meissner effect. Here, due to the finite size of the sample and the spacing between the diamond particles and the sample, there is a residue magnetic field which causes a Zeeman splitting of around 64~MHz for ${\rm NV_C}$ at low temperatures. There is also a difference in the Zeeman splitting for three diamond particles when the sample is in the normal state, because these diamond particles are randomly oriented relative to the applied magnetic field.

Fig. 2(h) demonstrates the major advantage of our technique. As discussed above, both the transverse and longitudinal components of the field relative to a given NV center can be calculated from its ODMR spectrum. This provides the means to reconstruct the field vector.
When the sample is in the normal state, the orientation of the NV center can be calibrated against the applied field direction, which is directed along the $c$-axis of the sample. This gives an effective magnetic field vector along the $c$-axis that can be tracked as a function of temperature (details in Supplementary).
With these considerations, we determine the effective magnetic field vector felt by ${\rm NV_C}$, ${\rm NV_E}$ and ${\rm NV_F}$ at 8.3~kbar. For ${\rm NV_C}$, the field vector shortens and tilts away from the vertical direction upon entering the superconducting state. This is consistent with the fact that ${\rm NV_C}$ is on the top of the sample, and that the Meissner effect causes the field lines to bend around the sample. However, for ${\rm NV_E}$, the field vector lengthens and it does not tilt much in the superconducting state, again consistent with the fact that ${\rm NV_E}$ is located off to the side of the sample, so that the field lines remain vertical but denser in the Meissner state there. Finally, the field vector sensed by ${\rm NV_F}$ remains practically constant across the superconducting phase transition, in stark contrast to the behavior of ${\rm NV_E}$ and ${\rm NV_C}$. The ability to collect the complete vectorial information with spatial resolution under extreme conditions represents one of the key advancements of our technique.


Next, we illustrate the performance of our setup under a varying pressure. In a separate run, we have calibrated our ODMR shift against the shift of the ruby fluorescence spectrum up to 60~kbar, thereby confirming our capability to sense the pressure (see Supplementary) and to conduct ODMR experiments at high pressures. It remains to show that our setup does not lose sensitivity to the superconducting transition when pressure is varied. Figure 3A displays the temperature dependence of the Zeeman splitting of ${\rm NV_C}$ at 7 different pressure points, from which the pressure dependence of $T_c$ can be detected. Additional supporting AC susceptibility data can be found in Supplementary. The resultant $T$-$p$ phase diagram is displayed in Fig. 3(b), showing a suppression of the superconducting state with pressure. This is consistent with the fact that $x=0.41$ is located at the overdoped side of the superconducting dome \cite{Lina10}. To verify the reproducibility, we also collect data on releasing pressure. The overall smooth evolution of $T_c$ against $p$ shows that the system is in the elastic regime. This series of experiments confirms the performance of our technique.

The transition width for both methods in Fig. 2(f) exhibits a noticeable difference. This is because, with the application of a magnetic field, the vortex state can be stabilized in a type II superconductor. The broadening of the ODMR splitting is due to the fact that the NV center located in the close proximity of the sample begins to sense the penetrating field in the form of vortices (AC susceptibility, which probes the average response of the whole sample, are much less sensitive to the vortex state). To probe the phase boundaries, we calculate the magnetic field along the sample $c$-axis sensed by ${\rm NV_C}$. Fig. 4(a) shows the temperature dependence of the resultant field at 8.3~kbar. 
Below $T_{c1}$ and above $T_{c2}$, the $c$-axis field is temperature independent. However, in between $T_{c1}$ and $T_{c2}$, a rapid rise of the $c$-axis field is detected. This is due to the entry of the magnetic field lines in the form of vortices at $T>T_{c1}$, and the full penetration of the applied magnetic field for $T>T_{c2}$. Using the data at 30~K, which is in the normal state, we can calibrate the value of the applied magnetic field. Thus, this magnetic field must be proportional to $H_{\rm c1}$ at $T_{c1}$, and equals to $H_{\rm c2}$ at $T_{c2}$. Hence, our ODMR data offer the exciting prospect of detecting the transition from the Meissner state to the vortex state under pressure.

\begin{figure}[!t]
\center
\includegraphics[width=0.5\textwidth]{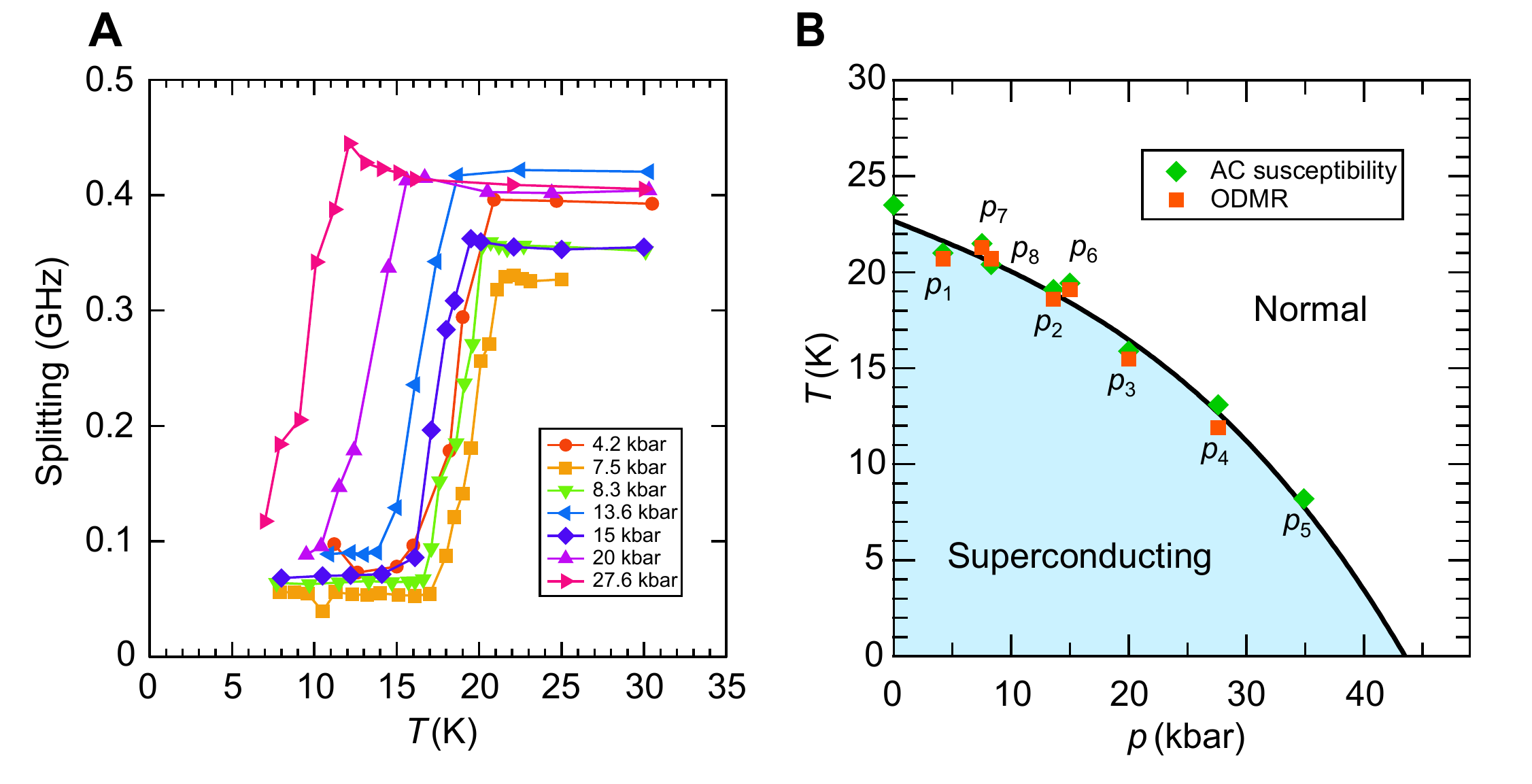}
	\caption{\textbf{Temperature-pressure phase diagram constructed using NV centers.}
   (\textbf{a}) The Meissner effect measured by the Zeeman splittings of NV centers under different pressures. The applied magnetic field is $(70 \pm 5)$~G. (\textbf{b}) The change of $T_c$, measured with ODMR method (green, diamond) and AC susceptibility (red, square), against the applied pressure. ``$p_1$...$p_8$" shows the sequence of the applied pressures. The error bars are smaller than the symbol size.}
\label{fig:3}
\end{figure}

\begin{figure}[!t]
\center
\includegraphics[width=0.5\textwidth]{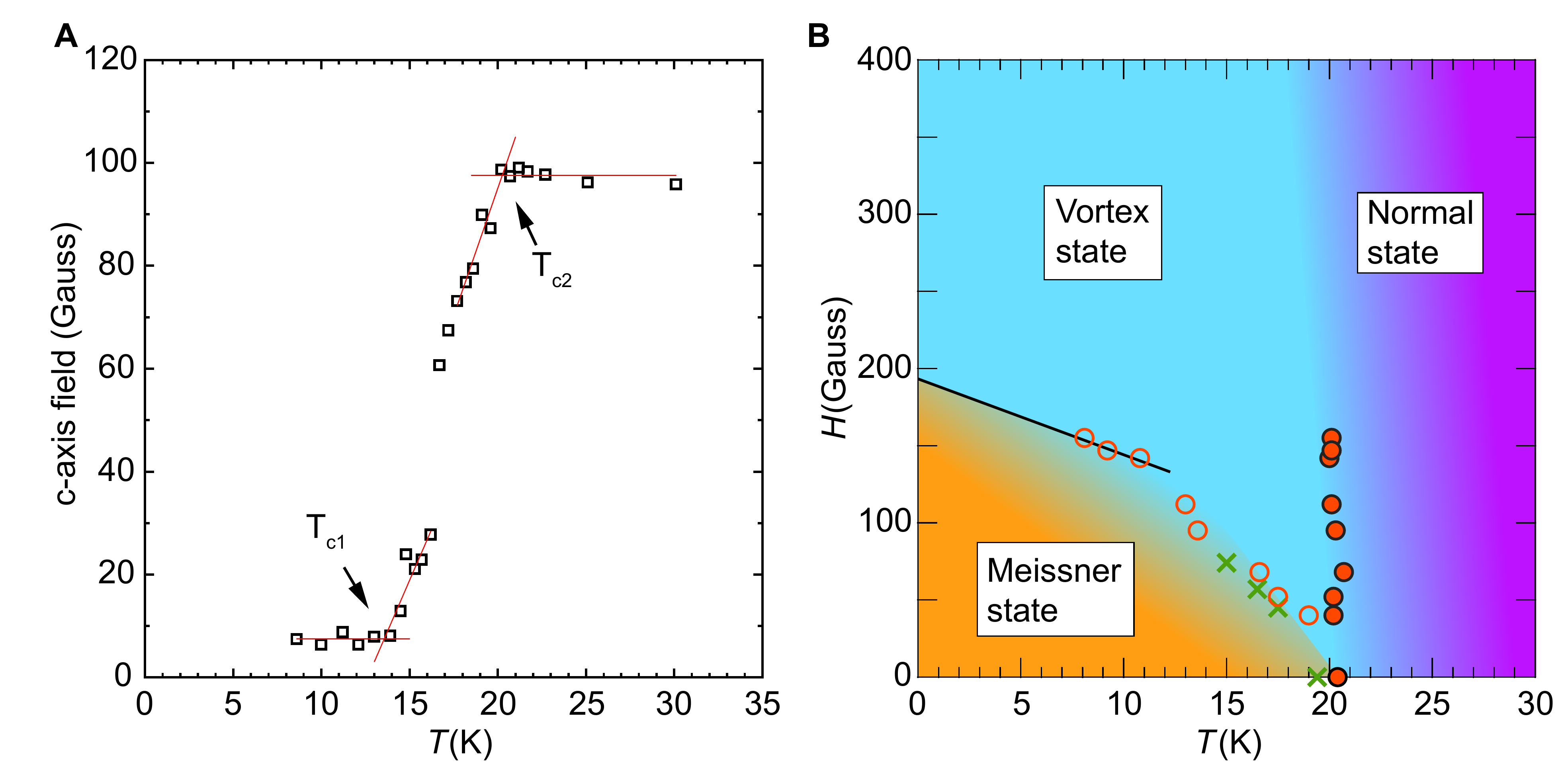}
	\caption{\textbf{Measurement of the lower critical field ($H_{\rm c1}(T)$) and the upper critical field ($H_{\rm c2}(T)$)}
   (\textbf{a}) The magnetic field along the $c$-axis measured for $\rm NV_C$. The applied field along the $c$-axis is 95 G, which can be determined from the data at 30~K. The definitions of $T_{c1}$ and $T_{c2}$ are shown. (\textbf{b}) Phase diagram showing $\alpha H_{\rm c1}(T)$ (red open circles) and $H_{\rm c2}(T)$ (red solid circles) at 8.3~kbar. Here, the geometry factor $\alpha$ for a thin slab with lengths $l_c$ along the field and $l_a$ perpendicular to the field can be calculated by $\alpha=\tanh\sqrt{0.36(l_c/l_a)}$ ~\cite{Brandt99}, where $l_c/l_a\sim 0.8$. Therefore, $\alpha$ is around 0.5. The dotted line is a guide to the eyes. Additional $\alpha H_{\rm c1}(T)$ for 15~kbar are added to the phase diagram for comparison (green crosses). The error bars are smaller than the symbol size.}
\label{fig:4}
\end{figure}


Repeating the measurements at different applied fields, we can trace out $\alpha H_{\rm c1}(T)$ and $H_{\rm c2}(T)$ for $x=0.41$ at 8.3~kbar (see Fig. 4(b)), where $H_{c1}(T)$ is the boundary between the Meissner state and the vortex state and $\alpha\sim0.5$ is a numerical constant that depends on the geometry of the sample. From $H_{\rm c1}(T)$, the temperature dependence of the London penetration depth can be deduced, allowing the discussion of the superconducting gap function \cite{Putzke2014,Auslaender15}. Note that $H_{\rm c1}(T)$ appears linear at low temperatures, and extrapolates to 384~G at 0~K. Both the linearity and the extrapolated $H_{\rm c1}(0)$ value are in good agreement with previous $H_{\rm c1}$ studies conducted for this family of Fe-based superconductors via micro-Hall probe array \cite{Putzke2014}. On the other hand, the initial slope $|dH_{\rm c2}/dT|_{T_c}$ is proportional to the square of the quasiparticle effective mass relative to the free electron mass, the almost vertical $H_{\rm c2}(T)$ is consistent with the strongly correlated nature of the material system.


In summary, we successfully demonstrate the usage of NV centers in diamond as a vector magnetic field sensor with superior spatial resolution and field sensitivity in pressure cells under cryogenic conditions. We demonstrate our sensing capability by studying superconductivity through Meissner effect. Using a piece of single-crystalline BaFe$_2$(As$_{0.59}$P$_{0.41}$)$_2$ as a model system, we construct the temperature-pressure phase diagram, and determine both the lower critical field and upper critical field. The spatial resolution of the protocol shown here can be pushed to sub-100~nm. This resolution will provide the unique opportunity to sense the dynamics of magnetically related features such as magnetic domains, vortices ~\cite{Vincent2013,Nini, Maletinsky2016, Folman18} and skyrmions in pressure cells.
As a non-invasive and contactless method, it can be used to study systems that are too small or too delicate for traditional macroscopic field sensors such as flakes of 2-D materials ~\cite{Cao2018}. What is more, not limited to magnetic field sensing, NV center is sensitive to other physical parameters such as local electric fields and mechanical strain too. Therefore, the method demonstrated here can be used beyond magnetic field related processes and becomes a powerful tool in studying quantum physics in strongly correlated systems.

\bibliography{references}

\end{document}